\documentclass[journal,12pt,a4paper]{IEEEtran}

\usepackage{amsmath}
\usepackage{amsfonts}
\usepackage{amssymb}
\usepackage{makeidx}
\usepackage{listings}
\usepackage{xcolor}
\usepackage{caption}
\usepackage{float}
\usepackage{hyperref}
\usepackage{comment}

\usepackage[english]{babel}

\usepackage{mathptmx}

\usepackage[utf8]{inputenc}

\usepackage{graphicx}
\graphicspath{ {images/} } 

\begin{document}

\title{Security analysis of a blockchain-based protocol for the certification of academic credentials}

\author{\IEEEauthorblockN{Marco Baldi, Franco Chiaraluce, Migelan Kodra and Luca Spalazzi}\\
\IEEEauthorblockA{Dipartimento di Ingegneria dell'Informazione\\
Universit\`a Politecnica delle Marche\\
Ancona, Italy, 60131\\
}}


\maketitle

\begin{abstract}
We consider a blockchain-based protocol for the certification of academic credentials named Blockcerts, which is currently used worldwide for validating digital certificates of competence compliant with the Open Badges standard.
We study the certification steps that are performed by the Blockcerts protocol to validate a certificate, and find that they are vulnerable to a certain type of impersonation attacks.
More in detail, authentication of the issuing institution is performed by retrieving an unauthenticated issuer profile online, and comparing some data reported there with those included in the issued certificate.
We show that, by fabricating a fake issuer profile and generating a suitably altered certificate, an attacker is able to impersonate a legitimate issuer and can produce certificates that cannot be distinguished from originals by the Blockcerts validation procedure. We also propose some possible countermeasures against an attack of this type, which require the use of a classic public key infrastructure or a decentralized identity system integrated with the Blockcerts protocol.
\end{abstract}



\section{Introduction}
The certification of competences and academic credentials plays a fundamental role in modern societies and everyday life. Classically, such a certification is performed through paper certificates containing seals and signatures. 
These documents, however, have no uniformity among different countries as well as no recognized digital equivalent and, most importantly, are subject to falsification.
This is confirmed by several cases occurred in recent years, as in the case of the dean of admissions at the Massachusetts Institute of Technology (MIT), who declared to have fabricated and lied about her own educational credentials for 28 years \cite{lewin_2017}.
In this context, an internationally recognized standard for the digitalization and authentication of competence certificates (as transcripts, diplomas, etc.) and academic credentials becomes a must.
Besides providing strong authentication mechanisms, such a system should also aim at establishing an internationally recognized format for portability and verification of competences.

The first step into verifying competences online and everywhere was made by the Mozilla Foundation \cite{mozillafoundation} in collaboration with the McArthur Foundation \cite{macarthurfoundation}, through the open-source project called \textit{Open Badges} \cite{openbadges}. The scope of Open Badges is to provide organizations and institutions with a system for issuing digital badges to the competence owners in order to recognize not only their official learning but even transversal skills. 
For this purpose, certificates compliant with Open Badges are designed to provide a detailed profile of the recipient, including the so-called \textit{Diploma Supplement} \cite{educationandtraining-europeancommission}, collecting all the academic records of a student, giving this way a clear picture about the competences gained during the academic path.
The Open Badges standard hence provides a tool for implementing digital, enriched versions of competence certificates and academic credentials.
However, the certification of the competences covered by one of these digital badges is out of the scope of the Open Badges standard itself.

With the aim to fill such a gap, the use of blockchain technology has emerged as a valuable solution for the validation of Open Badges-compliant certificates \cite{Mikroyannidis2018}.
The most widespread and internationally adopted blockchain-based system for the validation of these certificates was developed by the MIT Media Lab \cite{mitmedialab} in collaboration with Learning Machine \cite{learningmachine}, and is called \textit{Blockcerts} \cite{blockcerts}. 

According to the Blockcerts standard, Open Badges-compliant certificates are cryptographically signed by the issuer, while their certification data is written into a public blockchain, thus leveraging its immutability.
Recipients can share them publicly on their social media profiles, personal websites, etc., and everyone can see the contents of the certificates with the possibility to verify their validity and authenticity through the blockchain. 
The use of the blockchain technology makes the process of verification globally accessible and instantaneous, with no need for long procedures and institutional bureaucratic correspondences. 
Another important point lays in the fact that Blockcerts passes from paper-based certificates to software-based certificates. A Blockcerts certificate is fully-machine readable since it is designed as software in JavaScript Object Notation (JSON) format \cite{json}.

In this paper we review the steps of issuance and validation of a certificate compliant with the Blockcerts protocol, with the aim of analyzing its security and resistance to forgery.
In particular, we focus on the certificate authentication process through the blockchain, which is designed to be decentralized and self-consistent.
For this purpose, each certificate must contain all the necessary information for its validation through the blockchain, including a reference to the public key of the issuer to be used for its validation.
We show that such a feature, although allowing decentralized validation, opens the door to possible forgery attacks.
We first describe how an attack of this type can be mounted, and then we show its feasibility through a practical example.

For this purpose, we generate a forged but verifiable academic certificate. Such a forged certificate appears to be correctly issued by the \textit{Universit\`{a} Politecnica delle Marche} according to the Blockcerts protocol, despite it has been fabricated without involving the aforementioned institution.
In fact, by generating a valid blockchain key pair, and following a few simple steps, it was possible to impersonate the issuing institution and fabricate an apparently valid academic certificate.
This is due to the fact that the Blockcerts verification system does not check if the public keys are actually owned by the legitimate issuing institution or not.
In order to prevent this type of attacks, we propose some countermeasures aimed at avoiding that a fake issuer profile can be accepted by the certificate verification protocol.

The paper is organized as follows.
In Section \ref{sec:openbadgesblockcerts} we describe the Open Badges and Blockcerts standard that are the object of our work.
In Section \ref{sec:attack} we describe a vulnerability in the Blockcerts certificate validation procedure and how it can be exploited to fabricate forged academic credentials.
In Section \ref{sec:countermeasures} we describe some possible countermeasures aimed at preventing the aforementioned attack, while in Section \ref{sec:conclusion} we provide some conclusive remarks.

\section{Open Badges and Blockcerts\label{sec:openbadgesblockcerts}}
Next we briefly describe the Open Badges and Blockcerts standards, focusing on their features that are of interest for the purposes of our study.

\subsection{Open Badges}
Open Badges represents a group of specifications and open technical standards developed by the Mozilla Foundation \cite{mozillafoundation} with funding from the MacArthur Foundation \cite{macarthurfoundation} and a network of partners committed into developing a new way to recognize learning wherever it happened, in and out of formal education and online \cite{openbadges}.

In 2011, The Mozilla Foundation announced its plan to develop the technical standard called Open Badges as a response to the trend of documenting and representing skills and/or achievements online through badges. The Open Badges project defines an open specification and application program interfaces that provide any organization with the basic building blocks required to offer badges in a standard, interoperable manner \cite{surman_2011}. Open Badges also allow combining different badges together. 
This allows painting a complete profile of an individual's identity, achievements, hobbies, and skills, creating this way a digital passport of the relevant competences. 
Collecting all this information in a single badge makes the process of finding a job easier, or makes it easier to find the right person to hire for a job. Once this project started, there was a global interest and an international community was formed almost immediately.

In practice, an Open Badge is an image file that contains embedded metadata. 
Such data represent the scope of the badge, that is, they describe why it was awarded, who issued it, who earned it, etc. The metadata also contains information about the criteria to achieve the badge and how to verify the authenticity of that particular badge.  

According to the Open Badges standard \cite{openbadgesv2.0_2018}, data embedded into an Open Badge is collected into a JavaScript Object Notation - Linked Data (JSON-LD) \cite{json} context file \cite{openbadgesinfrastructurecontextfile}.
The core elements involved in the process of issuing an Open Badges-compliant digital badge are: i) Profile, ii) BadgeClass and iii) Assertion.
These three elements are combined to create the digital badge, and are described next, along with the links among them.

\subsubsection{Profile}

Every organization and individual issuing and receiving Open Badges is represented by a \textit{Profile} section \cite{openbadgesv2.0_2018}. The Profile section collects the information that identifies the issuer, thus providing an \textit{Issuer Profile}, or the receiver, thus providing a \textit{Receiver Profile}.
Concerning the Issuer Profile, the following information is mandatory \cite{openbadgesv2.0_2018}: i) id, ii) type, iii) name, iv) url and v) email.
The Issuer Profile may be completed with the following additional (optional) fields \cite{openbadgesv2.0_2018}: i) description, ii) image, iii) telephone, iv) public key, v) verification and vi) revocation list.
All these entries provide a full and clear picture of the issuer. 
Differently from the Issuer Profile, in the Recipient Profile the only mandatory fields are id and type \cite{openbadgesv2.0_2018}, while all the other elements are optional. 

\subsubsection{BadgeClass}

A BadgeClass contains the description of the achievement the Issuer recognizes \cite{openbadgesv2.0_2018}. As an example, for a student graduating at the university, a BadgeClass containing the information of the Degree must be created. The information embedded in BadgeClass contains fields such as the name and the description of the achieved degree, the image of the badge and the criteria on how this particular badge can be earned. In addition to this information, BadgeClass includes an Issuer Profile to identify the institution who created and issued the badge.
The full list of entries included in the BadgeClass is \cite{openbadgesv2.0_2018}: i) id, ii) type, iii) name, iv) description, v) image, vi) criteria and vii) issuer.
The list is completed with the following optional properties \cite{openbadgesv2.0_2018}: i) alignment, ii) tags.

\subsubsection{Assertion}

The collection of information representing the awarded badge plus the information about the recipient of the badge creates the so-called Assertion element \cite{openbadgesv2.0_2018}. The Assertion links to the BadgeClass and contains some specific information like the date in which the badge is awarded, the identifier of the badge recipient, and other additional information like an expiration date or an evidence url. The entries included in the Assertion section are \cite{openbadgesv2.0_2018}: i) id, ii) type, iii) recipient, iv) badge, v) verification, and vi) issuedOn, while the additional optional properties are \cite{openbadgesv2.0_2018}: i) image, ii) evidence, iii) narrative, iv) expires, v) revoked and vi) revocationReason.

\begin{figure}[H]
	\centering
	\includegraphics[width=1\linewidth]{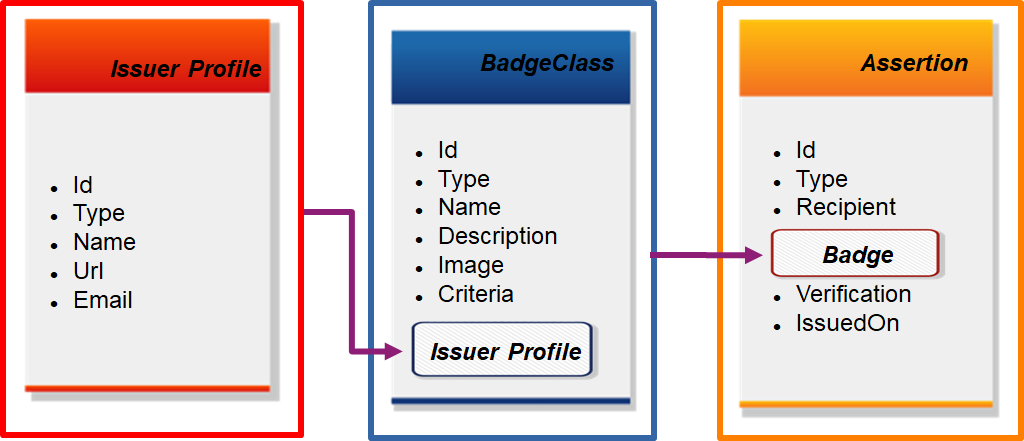}
	\caption{Open Badges issuing paradigm}
	\label{fig:obshema}
\end{figure}

The three sections Profile, BadgeClass and Assertion are combined together generating this way the Open Badge. A graphic view of the link between each section is shown in \autoref{fig:obshema}.

Based on the above elements, issuing of Open Badges involves the following steps:
\begin{enumerate}
\item Issuers create BadgeClasses and award assertions to the recipients. 
\item The recipient of an Open Badge is free to share it on social media, via e-mail or in some other form, with friends, family and potential employers. 
\item Everyone can check the authenticity of the badge and the fact that it was awarded to the person who claims it was awarded to through the information included in the badge itself. 
\end{enumerate}

\subsection{Blockcerts \label{sec:Blockcers}}

With the aim of having a global system for the verification of academic records, the development of Blockcerts \cite{blockcerts} was initiated in 2015 as part of a research project by the MIT Media Lab \cite{mitmedialab} in collaboration with Learning Machine \cite{learningmachine}. 
The Blockcerts project exploits the Open Badges framework jointly with the blockchain technology in order to realize a global, decentralized notary with the following claimed features \cite{learningmachine_2019}: i) Tamper evidence, ii) Issuer and recipient ownership, iii) Flexible form factor, iv) Online and offline sharing with verification and v) Independent verification.
The project was officially launched in 2016 and all the reference libraries were published under the MIT Open Source Licence, making the code accessible and free of charge. Therefore, Blockcerts is defined as an open standard for creating, issuing, viewing and verifying blockchain-based certificates.

Basically, the issuer can autonomously create a structured Blockcerts certificate, sign it and certify its integrity by storing a hash digest of the certificate within a blockchain transaction. 
Then, the issuer can send the recipient a copy of the signed Blockerts certificate that can be shared in social networks, via e-mail, etc.
Anyone accessing the certificate can verify its integrity using an open platform called \textit{Blockcerts Universal Verifier} \cite{blockcerts}, that performs a blockchain-based integrity check. 

\begin{figure}[ht]
	\centering
	\includegraphics[width=1\linewidth]{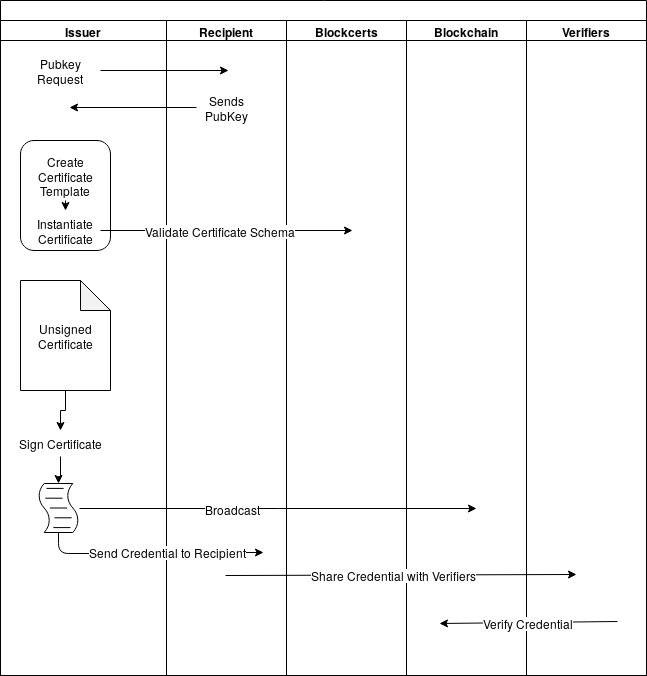}
	\caption{Sequence flow of the Blockcerts system}
	\label{fig:sequenceflow}
\end{figure}

The whole lifecycle of a Blockcerts-based certificate is schematically described in \autoref{fig:sequenceflow}.
After successfully completing her/his studies, the student is asked to provide a public blockchain address by the issuing institution.
For this purpose, the student generates a private/public keypair for the used blockchain, and then computes her/his public address as the output of a one-way function applied to her/his public key.
At the same time (or before) a Blockcerts-compliant certificate template is created by the issuing institution.
Then, a new certificate is issued and released to the student, along with a blockchain transaction to the student's public address that enables verification of the issued certificate.

More details about the Blockcerts standards, and the certification issuing and verification process are provided in the next sections.

\subsubsection{Blockcerts Certificate Design}

Each Blockcerts certificate contains structured data concerning the certificate itself, its issuer and its recipient.

Information about the certificate concerns:
\begin{enumerate}
\item certificate description,
\item certificate title,
\item certificate criteria,
\item embedded images,
\item certificate template.
\end{enumerate}

Information about the issuer concerns:
\begin{enumerate}
    \item \textbf{issuer\_url}: URL of the issuing institution website.
    \item \textbf{issuer\_email}: email address of the issuing institution.
    \item \textbf{issuer\_name}: name of the issuing institution.
    \item \textbf{issuer\_id}: URL where the \textit{Hosted Issuer Profile} can be found.
    \item \textbf{revocation\_list}: URL of the list of revoked certificates.
    \item \textbf{issuer\_signature\_lines}: information about who issued a Blockcerts certificate. This field is introduced considering the fact that in big organizations, there are different entities/persons/departments that can issue certificates.
    \item \textbf{issuer\_public\_key}: issuer blockchain public key. For the sake of correctness, we observe that, although being denoted as public key, this indeed is a public identifier obtained as the output of a one-way function applied to the real public key, as it occurs in the Bitcoin and Ethereum blockchains.
\end{enumerate}

\begin{figure}[H]
	\centering
	\includegraphics[width=1\linewidth]{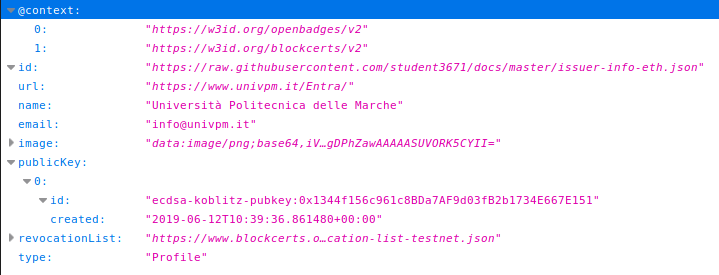}
	\caption{Hosted issuer profile}
	\label{fig:urlid}
\end{figure}

For the purpose of this work, an important field among those above is the issuer id, which is the URL of a web page containing the issuer information in JSON format. An example is shown in \autoref{fig:urlid}.

As we see from \autoref{fig:urlid}, such information includes the institution name, homepage, logo, e-mail address, etc., and, importantly, the public keys claimed by the issuer. Each key has a timestamp corresponding to its creation and possibly the timestamp corresponding to its expiration.

Information about the recipient instead concerns:
\begin{enumerate}
\item name,
\item email,
\item blockchain public key.
\end{enumerate}

In addition to the aforementioned information, the Blockcerts standard allows embedding the so-called \textit{diploma supplement} into a certificate.
The diploma supplement is produced by higher education institutions according to international and European recommendations, and includes the following eight sections:

\begin{enumerate}
    \item Information identifying the holder of the qualification.
    \item Information identifying the qualification.
    \item Information on the level of the qualification.
    \item Information on the contents and results gained.
    \item Information on the function of the qualification.
    \item Additional information.
    \item Information identifying the issuer of the qualification.
    \item Information on the national higher education system.
\end{enumerate}

It includes additional information about the course attended (e.g., results for each exam/course with date and outcome) and the competences achieved by the recipient of the certificate.
The diploma supplement is recognized as an official document accompanying a higher education diploma, providing a standardized description of the nature, level, context, content and status of the studies completed by its holder. 

\subsubsection{Issuing Blockcerts Certificate}

All the preliminary steps required for issuing certificates compliant with the Blockcerts standard can be implemented through the \textit{cert-tools} open source software \cite{github-cert-tools}, which allows generating a certificate template first, and then instantiating such a template into one or more certificates.
This allows generating an unsigned certificate, which then has to be signed and written into the blockchain network, besides sending a copy to the recipient.
These steps can be performed through the  \textit{cert-issuer} software tool \cite{github-cert-issuer}.
Delivery of the certificate is performed by creating a transaction from the issuing institution to the certificate recipient on the Bitcoin or Ethereum blockchain that includes the hash of the certificate itself. While it is possible to issue one certificate with one Bitcoin/Ethereum transaction, a solution for reducing the amount of data written to the blockchain is that of using one Bitcoin/Ethereum transaction to issue a batch of certificates.
This is possible through the generation of a Merkle tree of certificate hashes, in such a way that only the Merkle tree root has to be written into the blockchain.

\subsubsection{Certificate Authenticity Verification}
\label{sub:certificateauthenticity}
Through the verification process, anyone can verify the authenticity of a certificate, having clear information about the institution that issued it and a proof that the certificate was actually issued to the claiming recipient.
The certificate verification process starts with the following three steps:
\begin{enumerate}
    \item Verification that the hash digest of the certificate matches the value in the receipt.
    \item Verification that the Merkle path is valid.
    \item Verification that the Merkle root stored into the blockchain matches the value in the receipt.
\end{enumerate}

\begin{figure}[H]
	\centering
	\includegraphics[width=1\linewidth]{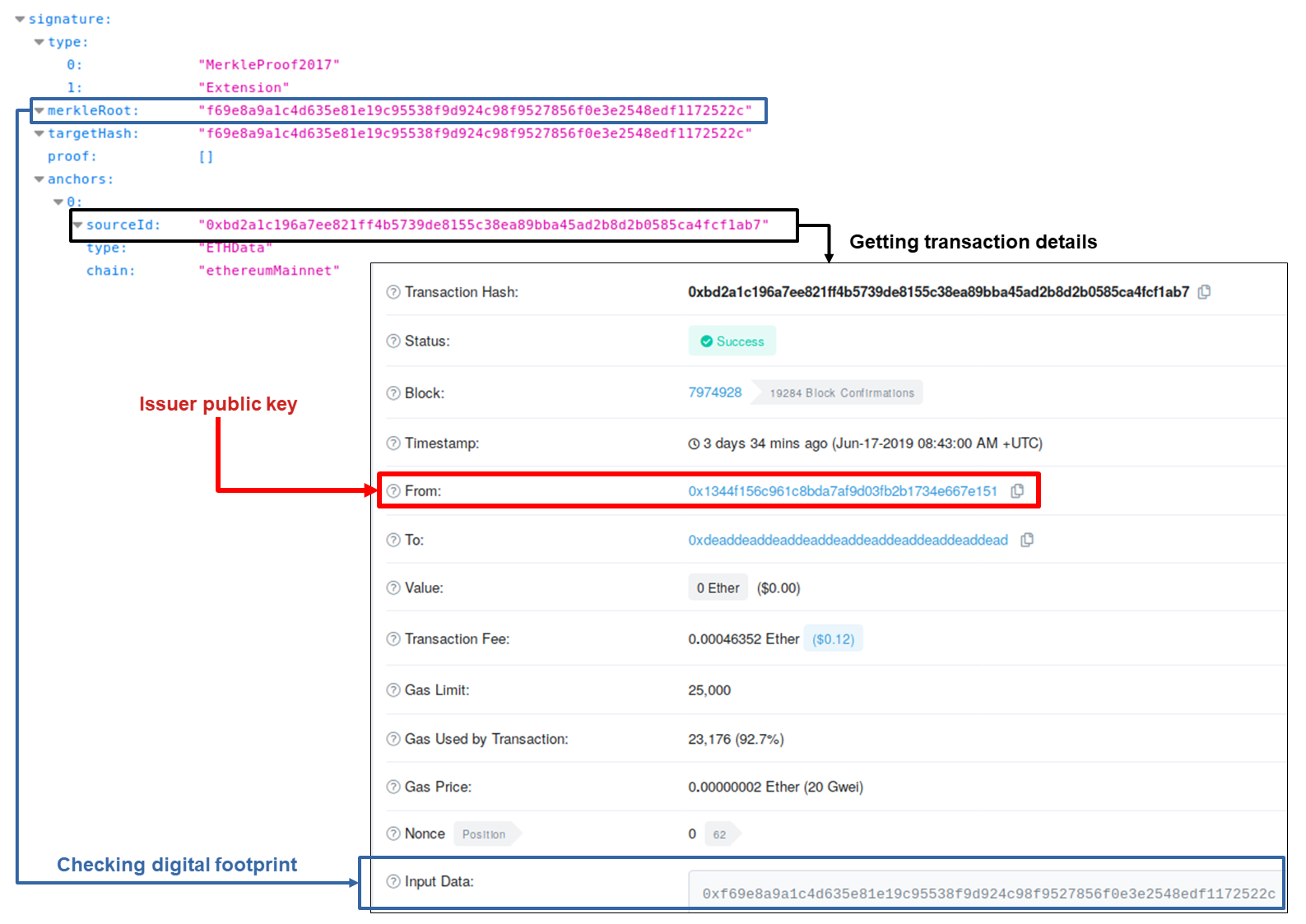}
	\caption{Extraction of the issuer public key from the transaction details}
	\label{fig:issuer-id-verif}
\end{figure}

Through the above steps, anyone can verify that a certificate has not been tampered since its issuing.
The next important step of the verification process is authenticating the certificate issuer, i.e., verifying the identity of the issuing institution. 
This is achieved by verifying that the signing key for the blockchain transaction through which the certificate was issued corresponds to the issuer public key, and that it was valid when the transaction took place. This uses the timestamp and input address from the blockchain transaction details, and the issuer information provided along with the \textit{Issuer Profile}, described in Section \ref{sec:Blockcers}. 
For this purpose, the blockchain transaction id is extracted from the certificate receipt. 
The transaction id allows verifying that the transaction was actually registered into the blockchain and retrieving the corresponding transaction details. The issuer public key can be found within such details, as shown in \autoref{fig:issuer-id-verif}.

\begin{figure}[H]
	\centering
	\includegraphics[width=1\linewidth]{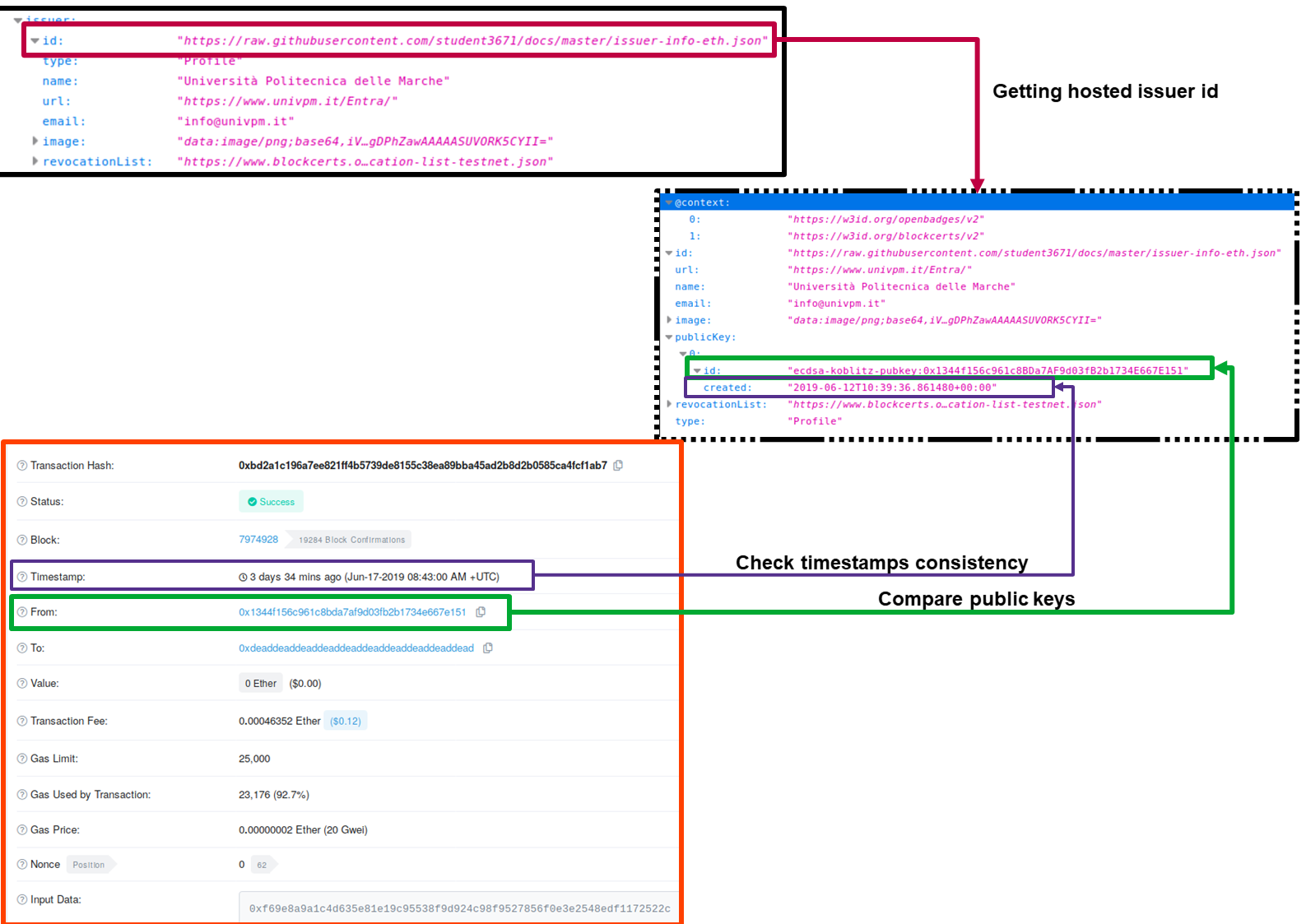}
	\caption{Issuer identity verification}
	\label{fig:issuer-id-host-verif}
\end{figure}

Then, the \textit{issuer\_id} is extracted from the certificate to retrieve the \textit{Hosted Issuer Profile} from the corresponding url.
The issuer public key included in the \textit{Hosted Issuer Profile} is then compared with the one found in the blockchain transaction, as shown in \autoref{fig:issuer-id-host-verif}.

\begin{figure}[H]
	\centering
	\includegraphics[width=1\linewidth]{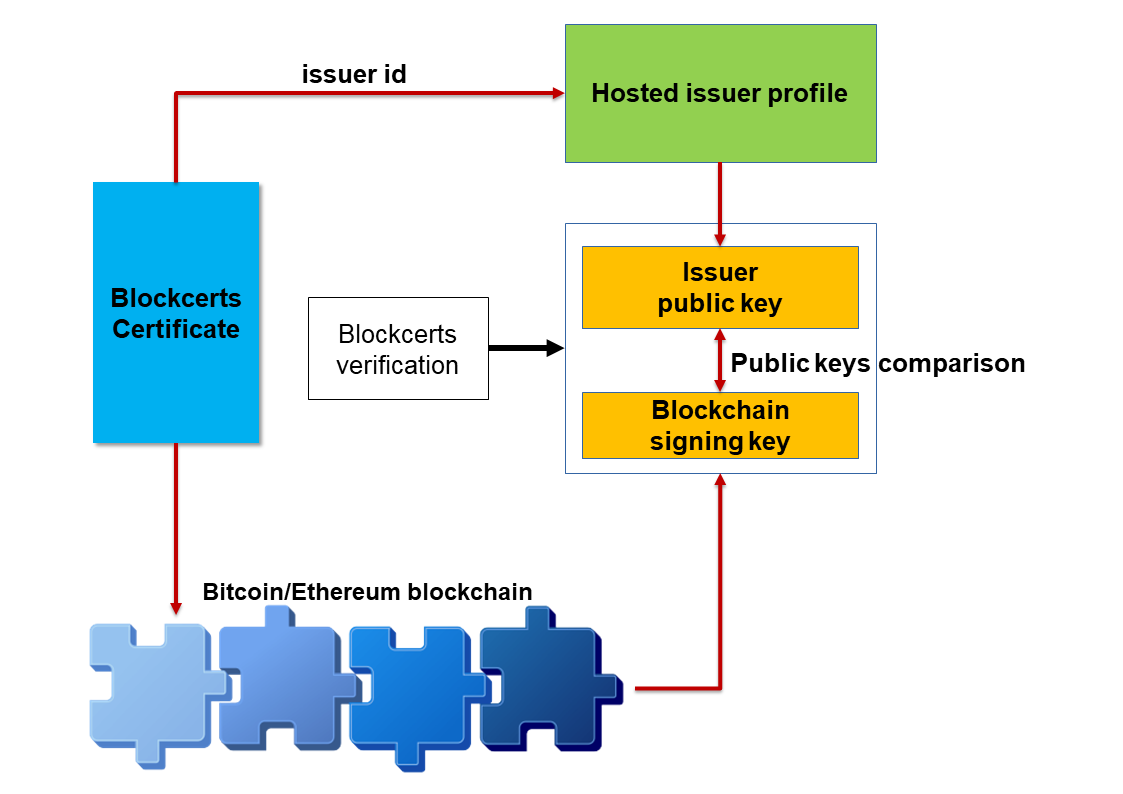}
	\caption{Blockcerts issuer verification process}
	\label{fig:verificationprocess}
\end{figure}

If the two keys do not coincide, an error is returned and the certificate is considered invalid. Otherwise, the timestamps are checked to prove that the key was valid at the time of the transaction. 
In addition, if the public key included in the issuer profile has an expiration date, it is checked that the transaction did not take place after that date.
If all these verification steps succeed, then the certificate is considered as valid.
All the steps of the issuer identity verification are schematically described in \autoref{fig:verificationprocess}.

\section{Fabricating academic credentials\label{sec:attack}}
In this section we show that apparently valid certificates issued by any institution can indeed be fabricated by a malicious attacker.
This is basically due to an intrinsic vulnerability of the verification process described in Section \ref{sub:certificateauthenticity}.
In fact, the Blockcerts protocol does not verify that the \textit{issuer\_id} extracted from a certificate indeed points to a web address that is owned by the legitimate issuing institution.
This allows hijacking of the verifier towards a fake issuer profile, which can perfectly resemble the one of the legitimate institution.  

Actually, a fake issuer profile was created for the Universit\`{a} Politecnica delle Marche and hosted on a Github domain. The public key included in such a profile has no relation with the real institution, since it was auto-generated for the purposes of this work. During the verification process, the Blockcerts protocol  checks the public key on the blockchain transaction corresponding to the certificate, and compares it with the key included in the issuer profile published online.
As we show next, this brings to a successfull verification through Blockcerts, and to a forged certificate that is practically indistinguishable from a legitimate one.
This results in the fabrication of a certificate for the Italian Laurea in Electronic Engineering by impersonating the issuing institution Universit\`{a} Politecnica delle Marche.

\subsection{Creation of the certificate}

Two Ethereum blockchain keypairs were first generated: one for the issuer (university) and one for the recipient (student).
After that, a Blockcerts-compliant certificate has been designed.
This step can obviously be skipped if the issuer already has defined its own Blockcerts-compliant certificates.

\begin{figure}
	\centering
	\includegraphics[width=0.92\linewidth]{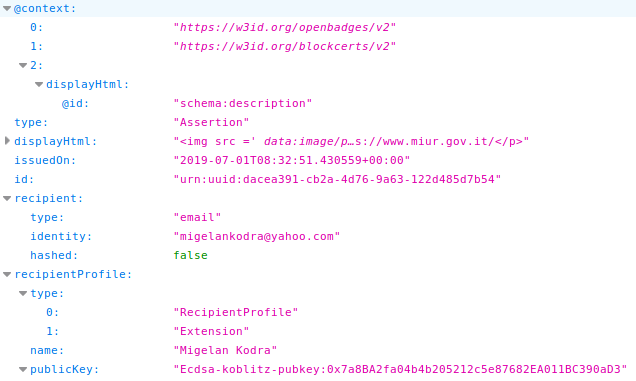}
	\includegraphics[width=0.92\linewidth]{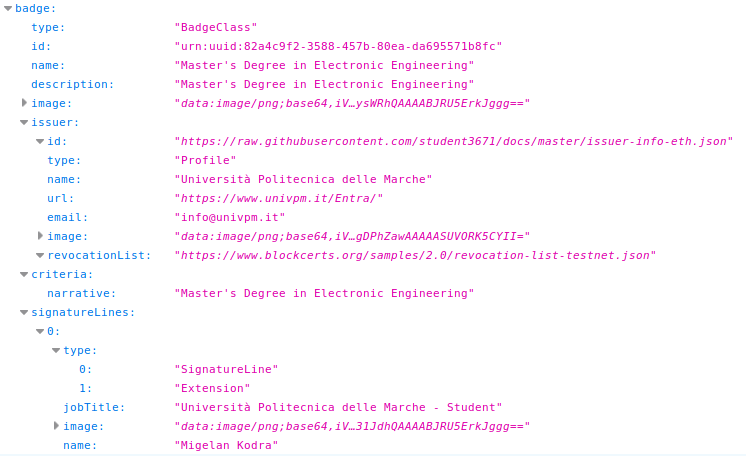}
	\includegraphics[width=0.92\linewidth]{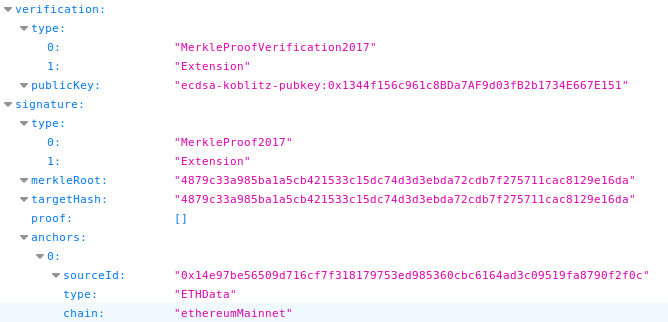}
	\caption{Blockcerts signed certificate}
	\label{fig:signedcertificate}
\end{figure}

A new certificate is then issued to the student, including the blockchain public key of the student, the student's personal information and a diploma supplement.
The correctness of this information is validated according to the Blockcerts and Open Badges standards, after which an unsigned certificate is obtained.
Such a certificate is then signed with the issuer keypair and issued through a transaction on the Ethereum blockchain.
The main content of such a blockchain transaction is the certificate Merkle root. In our case, since one single certificate was issued, the Merkle root coincides with the certificate hash digest. 
This way, a signed Blockcerts-compliant certificate is obtained, as shown in \autoref{fig:signedcertificate}.
Note that the signed certificate includes the blockchain receipt, providing all the necessary details for retrieving the blockchain transaction and performing the blockchain-based verification of the certificate according to the Blockcerts standard.

\begin{figure}
	\centering
	\includegraphics[width=1\linewidth]{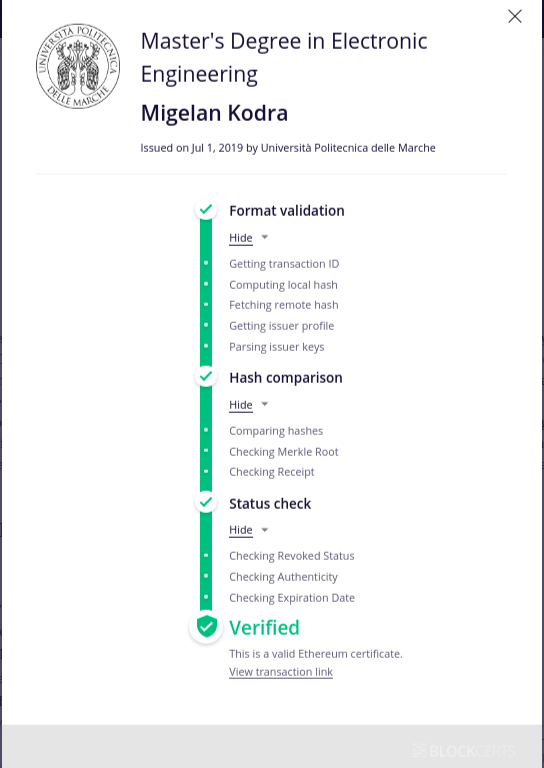}
	\caption{Certificate verification through the Blockcerts Universal Verifier}
	\label{fig:verifiedcert}
\end{figure}

This signed certificate can be verified through any web-based or stand-alone tool compliant with the Blockcerts protocol.
Let us use for this purpose the Blockcerts Universal Verifier\footnote{\url{https://www.blockcerts.org/}}. The outcome of the verification, so performed, is shown in \autoref{fig:verifiedcert}. The verification process indeed ended with a positive result.
Moreover, the verification window reports the issuing university logo and the title of the certificate (in this case: \textit{Master's Degree in Electronic Engineering}), followed by the name of the recipient, the issuance date and the name of the issuing institution.
The diploma supplement and its contents can be visualized as well.

\begin{figure}
	\centering
	\includegraphics[width=1\linewidth]{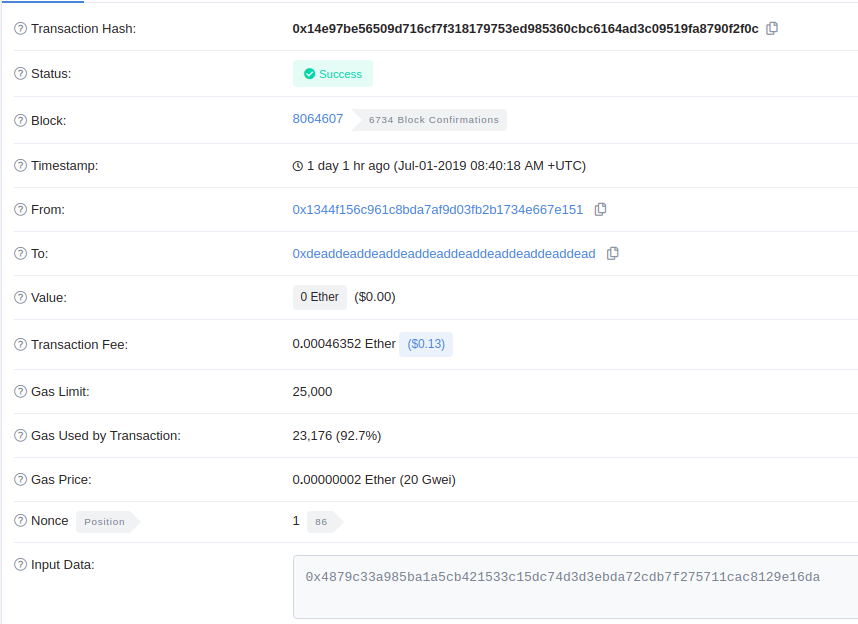}
	\caption{Ethereum blockchain transaction details}
	\label{fig:tx_details}
\end{figure}

The details of the corresponding Ethereum transaction\footnote{Available at
https://etherscan.io/tx/0x14e97be56509d716cf7f318 179753ed985360cbc6164ad3c09519fa8790f2f0c
} 
are reported in \autoref{fig:tx_details}.

\subsection{Analysis of the verification process}

Let us describe how it is possible that a fabricated certificate passes all the verification steps required by the Blockcerts protocol by analyzing them in detail.

As shown in \autoref{fig:verifiedcert}, the first verification step consists in reading the transaction id, through which the details reported in \autoref{fig:tx_details} are retrieved.
Then, a local hash digest of the certificate is computed, and the remote hash digest is retrieved from the associated blockchain transaction.

\begin{figure}[H]
	\centering
	\includegraphics[width=1\linewidth]{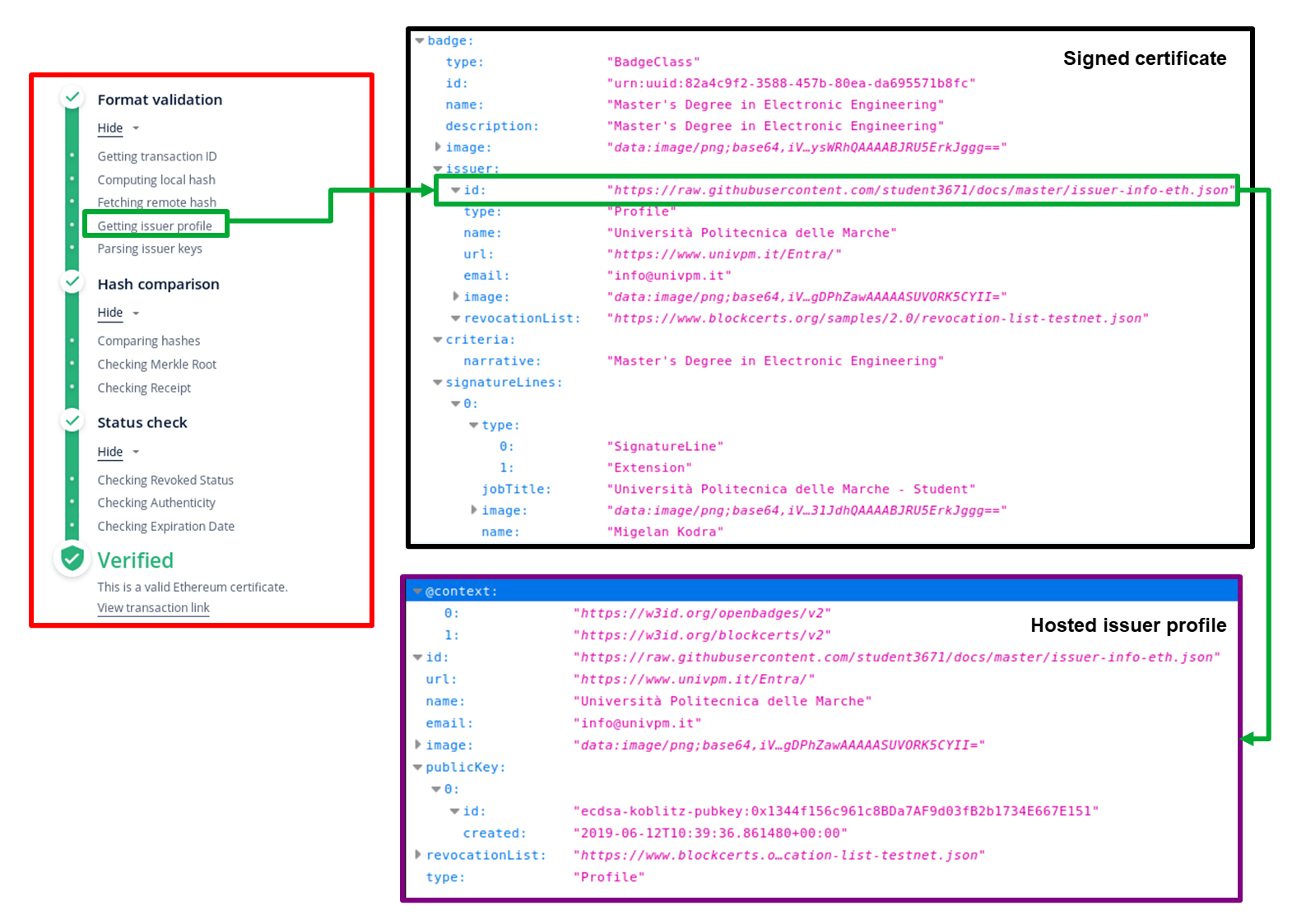}
	\caption{Getting the issuer profile}
	\label{fig:get_issuer_profile}
\end{figure}

The next verification step consists in getting the issuer profile referenced by the issuer\_id.
This is a crucial point for the successfull verification of a fabricated certificate, and is schematically described in \autoref{fig:get_issuer_profile}.
As we notice from the figure, the issuer\_id points to a custom url that has no relation with the legitimate issuer. This is the point where hijacking of the verification occurs, and the use of a fabricated issuer profile is enforced.
The next step consists in parsing the issuer keys, that is, extracting the issuer public key from the public  issuer profile, which in our case was fabricated.

The verification process continues with a second block of steps concerning the hash digest comparison. 
The first of these steps verifies that the hash of the certificate locally computed coincides with that included in the certificate receipt.
This proves that the certificate has not been modified since its issuance.
In the next step, the system compares the Merkle root value written on the certificate receipt with the Merkle root written on the blockchain. 
As a next step, the certificate receipt is checked to verify that the certificate under analysis is part of the Merkle tree.
In our case, only one certificate was issued and its hash digest corresponds with the Merkle root value.
This is denoted by leaving the \textit{proof} field empty in the certificate receipt. In such a case, the system realizes that only one certificate was issued and verifies that the Merkle root values in the certificate and blockchain are correct and coincide with the certificate hash digest.

\begin{figure}[H]
	\centering
	\includegraphics[width=1\linewidth]{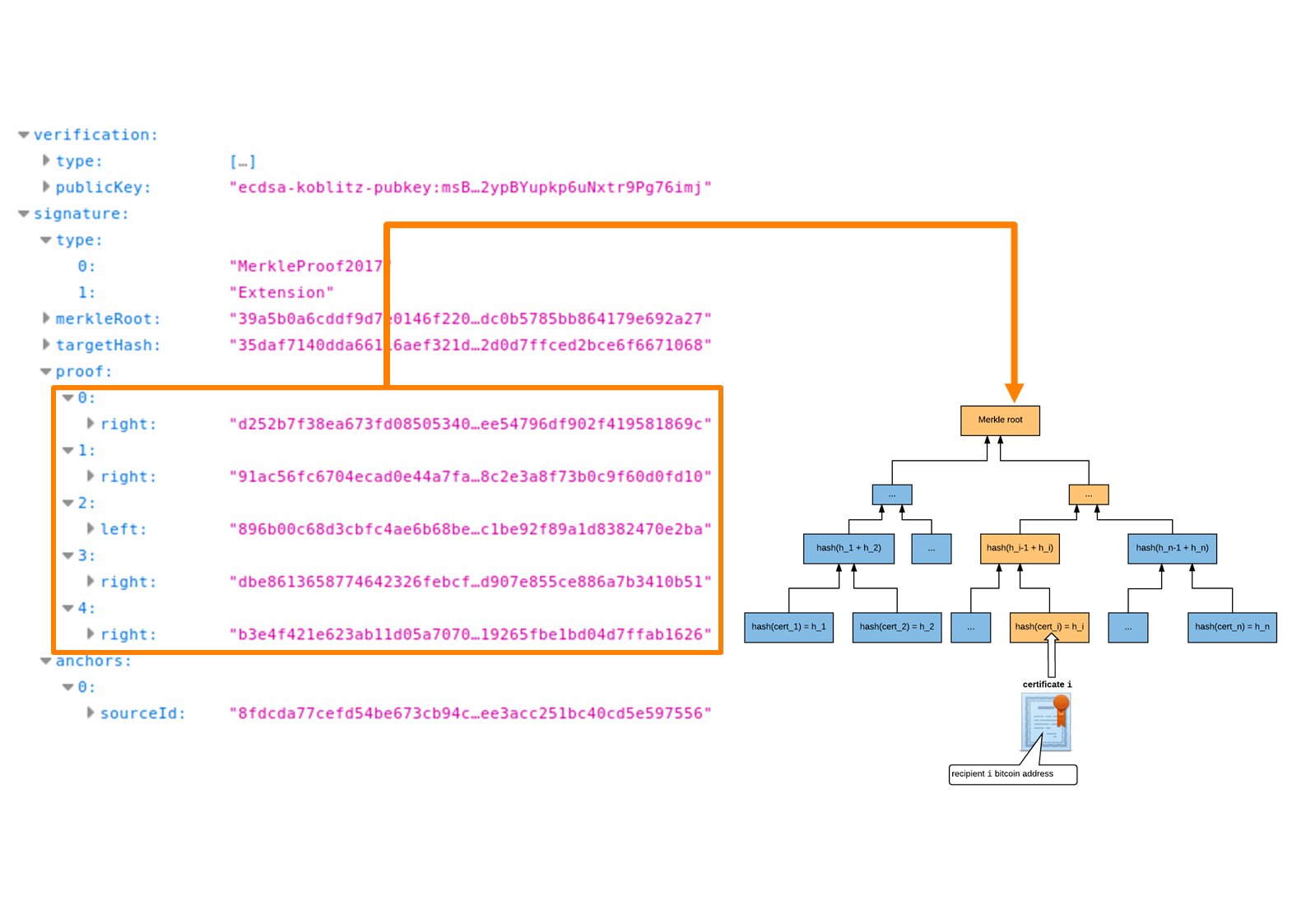}
	\caption{Merkle tree path for the verification of a certificate}
	\label{fig:receipt_proof}
\end{figure}

Instead, when a batch of certificates is issued, the \textit{proof} field is filled with the necessary values to create the path from the current certificate to the Merkle root. An illustrative example is shown in \autoref{fig:receipt_proof}, in which the Merkle path is colored in orange.
With this information, the system is able to calculate the final value of the Merkle root of the whole batch of certificates. If the certificate is part of the batch, then the calculated Merkle root value is the same as the value written on the certificate.
Otherwise, an error occurs, meaning that the certificate is not part of the batch, and therefore, is not valid.

\begin{figure}[H]
	\centering
	\includegraphics[width=1\linewidth]{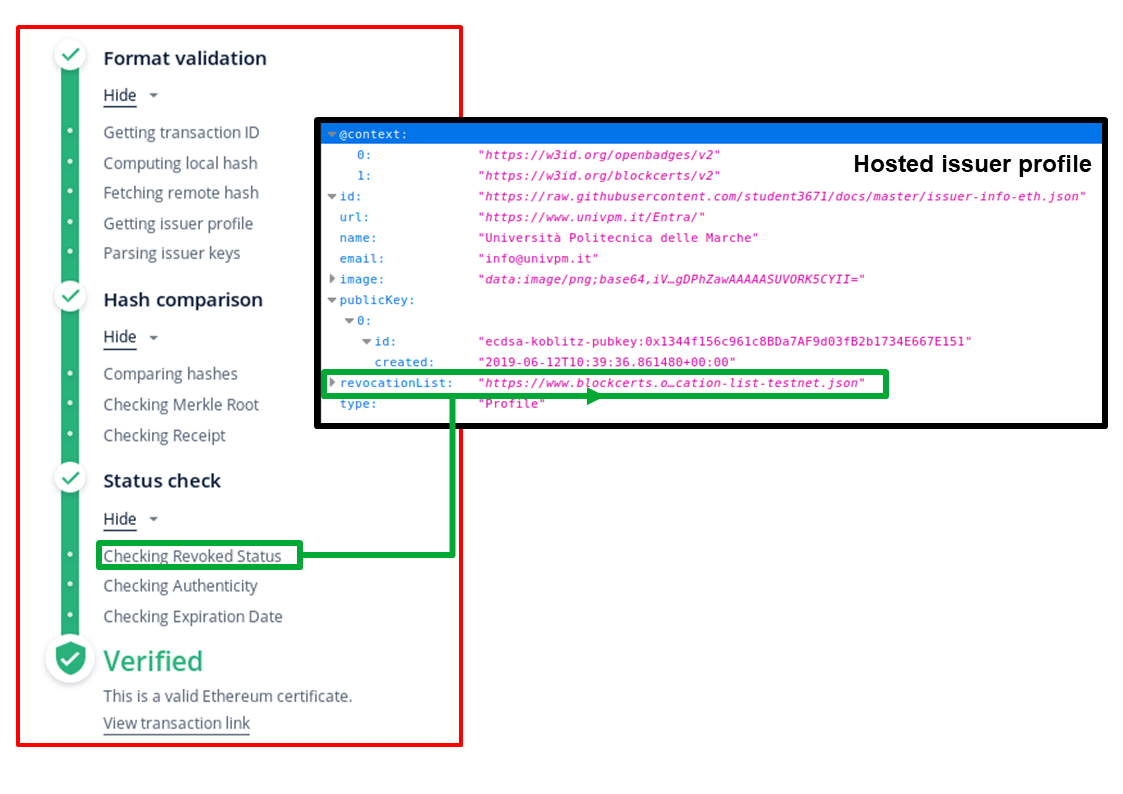}
	\caption{Checking revocation of a certificate}
	\label{fig:revocation_check}
\end{figure}

At this point, it has been proved that the certificate has not been modified since its issuance and that it is actually written on the blockchain.
The next block of verification steps, named \textit{status check}, concerns the certificate authentication.
The first one of these checks concerns the 
revocation status, and is aimed at verifying that the certificate has not been revoked by the issuer. 
In fact, a list of revoked certificates is available online along with the issuer profile, as shown in \autoref{fig:revocation_check}.
The system goes through such a list to check if it includes the certificate under analysis or not. In case the certificate identifier is found in the list of revoked certificates, an error message is returned and the verification fails, otherwise the system continues with the next verification step.

\begin{figure*}[ht]
	\centering
	\includegraphics[width=1\linewidth]{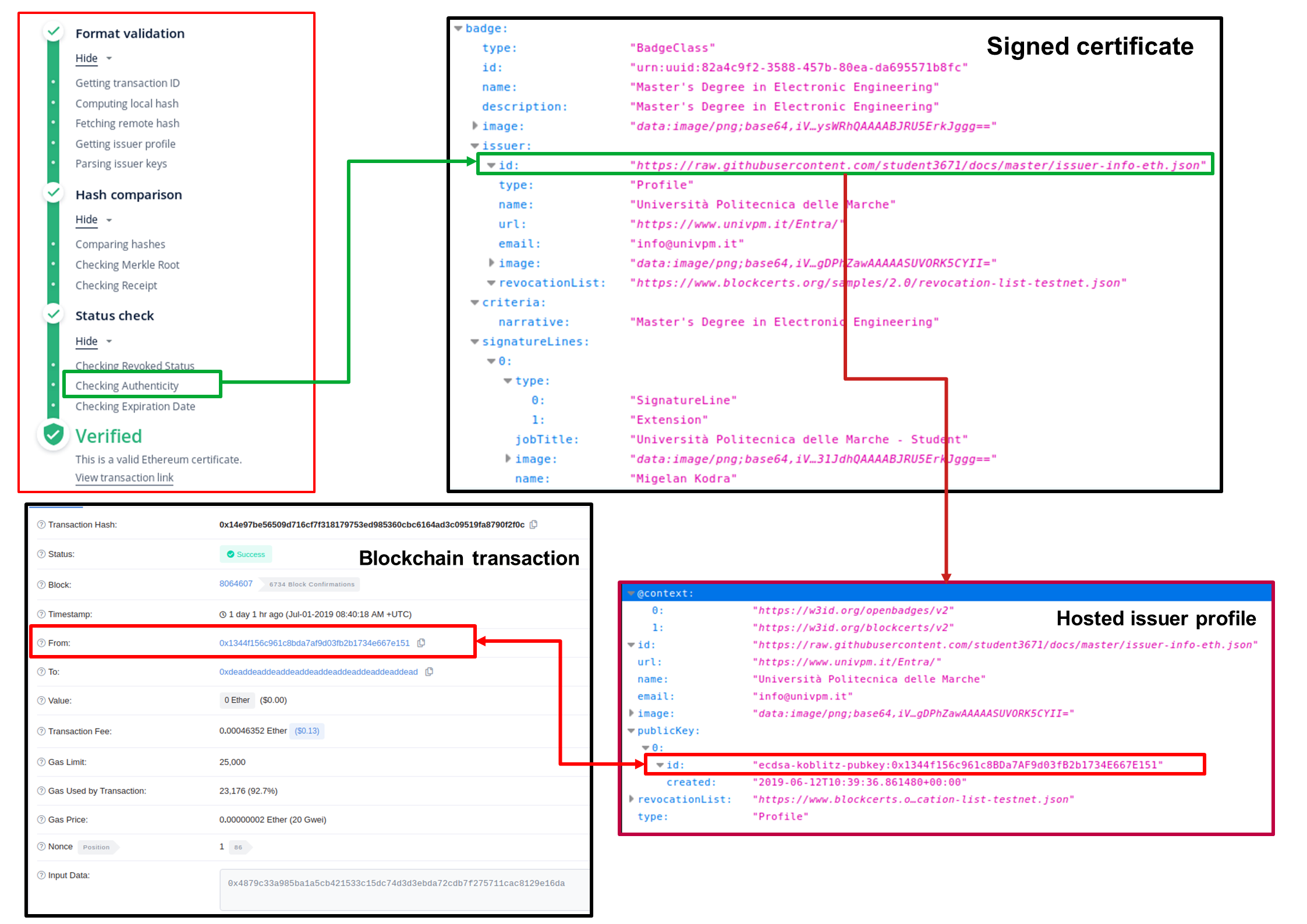}
	\caption{Checking authenticity of a certificate}
	\label{fig:checking_authenticity}
\end{figure*}

The subsequent step, named authenticity checking step, is a very crucial element in the verification process. This step aims at verifying that the certificate was actually issued by the claimed institution.
According to the Blockcerts standard, the certificate authenticity is checked as explained in Section \ref{sub:certificateauthenticity}. 
For the certificate under analysis, the steps performed for checking its authenticity are schematically described in \autoref{fig:checking_authenticity}.

For this purpose, the hosted\_id field is first read from the certificate, and the corresponding public key is retrieved from the issuer profile available online, at the web address specified in the hosted\_id field.
Such a public key is then compared with the one reported in the blockchain transaction. If the two keys coincide, the system checks the timestamps as explained before.
If such a check is successful, verification proceeds by checking if the certificate has an expiry date, after which the verification process is completed.

In our case, the fabricated certificate was able to pass all the verification steps, and is therefore considered as a valid Blockcerts-compliant certificate.
This proves that such a protocol does not allow distinguishing the legitimate issuer from someone impersonating it.
This is due to the fact that the Blockcerts standard does not require any verification that the keys used to sign a certificate are actually owned by the legitimate issuing institution.
For this reason, everyone creating a new keypair can sign a Blockcerts-compliant certificate and impersonate the legitimate issuing institution.

\section{Possible countermeasures\label{sec:countermeasures}}
The vulnerability of the Blockcerts standard described in the previous sections builds upon the lack of any certification about the ownership of the keys used for signing issued certificates.
In order to prevent forgery attacks exploiting such a vulnerability, suitable countermeasures must be adopted by introducing a mechanism to check that such keys indeed correspond to the digital identity of the legitimate issuing institution.

A natural solution of this type could be replacing the issuer profile referenced from the \textit{issuer\_id} field with a digital certificate containing the public key of the issuing institution, and released by an accredited certification authority.
In this way, the authenticity of the certificate can be checked through a classic Public Key Infrastructure (PKI), and the online information about the issuer retrieved through the \textit{issuer\_id} is certified.

\begin{figure}[ht]
	\centering
	\includegraphics[width=1\linewidth]{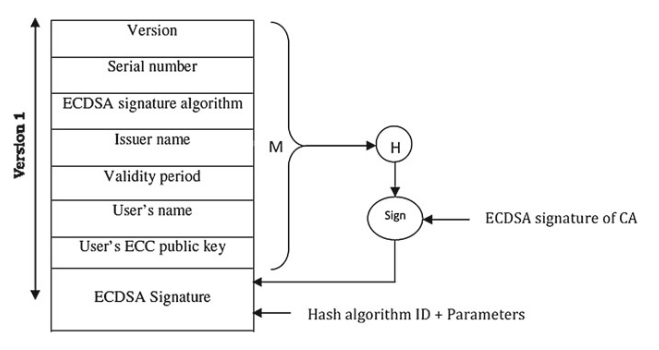}
	\caption{ECC-based X.509 certificate format}
	\label{fig:ecc_cert_format}
\end{figure}

Starting from July 2016, the eIDAS (electronic identification, authentication and trust services) came into effect in the European Union. The eIDAS regulation defines three types of electronic signatures \cite{eidas-regulation}:
\begin{enumerate}
\item Electronic Signature (data in electronic form which is attached to or logically associated with other data in electronic form and which is used by the signatory to sign).
\item Advanced Electronic Signature (an electronic signature which meets the following requirements: a) it is uniquely linked to the signatory, b) it is capable of identifying the signatory, c) it is created using electronic signature creation data that the signatory can use, with a high level of confidence, under his sole control, and d) it is linked to the data signed therewith in such a way that any subsequent change in the data is detectable).
\item Qualified Electronic Signature (an advanced electronic seal, which is created by a qualified electronic seal creation device, and that is based on a qualified certificate for electronic seal). \end{enumerate}
The only type of signatures universally recognized by all EU states and given the equivalent legal effect of a handwritten signature are qualified signatures.
The eIDAS regulation also provides the requirements for qualified certificates. 
Since most of the blockchain platforms (like the Ethereum blockchain used in our case) use ECDSA signatures, a natural solution would be using an ECC-based X.509 certificate format to link the issuer identity with its ECC-based public key.
The contents of a certificate of this type are schematically represented in \autoref{fig:ecc_cert_format} \cite{ecc_cert}.

This solution provides an effective way to overcome the Blockcerts issuer identity verification problem and to give a legal value to the Blockcerts academic certificate, owing to eIDAS compliance.
On the other hand, the ECC-based certificate is signed by a Trusted Service Provider (TSP), and validity of the certificate relies on validity of the TSP signature. 
The main drawback of such a solution is in the fact that, this way, the verification system is no longer decentralized, needing a Certificate Authority (CA) to verify the identity of the issuer and sign the ECC-based certificate. 

In order to overcome such a drawback, Decentralized Identifiers (DIDs) could be considered as an alternative solution.
In fact, there are currently working groups in the W3C and in the Decentralized Identity Foundation (DIF) with the goal to achieve a common understanding of the general architecture of decentralized identity systems based on Blockchain and DLT (Distributed Ledger Technology), and to develop standards that enable interoperability between different implementations even on different DLT platforms, while following privacy and security-by-design principles \cite{cen/cenelecfocusgroupbdlt_2018}. 
Some of these initiatives are \cite{cen/cenelecfocusgroupbdlt_2018}:

\begin{enumerate}
    \item W3C Community Group - Decentralized Identifier
    \item W3C Working Group - Verifiable Claims
    \item DIF - DID Auth
\end{enumerate}

Although these initiatives still do not provide consolidated standards and practices, in the mid term they are expected to represent an effective solution to restore the completely decentralized nature of Blockcerts and similar protocols while countering forgery attacks like those described in this paper.

\section{Conclusion\label{sec:conclusion}}
We have analyzed a blockchain-based protocol for the certification of academic credentials named Blockcerts, which aims at certifying digital certificates compliant with the Open Badges standard through a public blockchain.
We have reviewed all the steps that are required for the creation of Open Badges-compliant digital academic credentials and for their certification according to the Blockcerts protocol.
From such an analysis it results that the Blockchain protocol does not provide any strong mechanism for authenticating the issuing institution, since the issuer authentication is basically performed on the basis of an unauthenticated issuer profile available online and referenced from inside the certificate.

We have shown how a legitimate issuing institution can be easily impersonated by suitably fabricating a fake issuer profile.
This way, apparently legitimate academic credentials can be released, which the Blockcerts validation mechanisms are unable to distinguish from valid academic credentials issued by the legitimate institution.
This clearly highlights a vulnerability of this protocol, especially when it is used for the certification of academic credentials with legal value.

In order to overcome such a vulnerability, we have proposed to resort to a classic PKI and replace the issuer profile with a certificate signed by a recognized certification authority.
This, however, infringes the decentralized nature of the Blockcerts infrastructure.
Alternatively, a decentralized identity system could be used instead of a classic PKI to preserve the fully decentralized nature of the paradigm.
Such systems, however, are currently under development, and therefore cannot provide an immediate solution to the highlighted vulnerability.

\bibliographystyle{IEEEtran}
\bibliography{bibarchive}

\end{document}